\documentclass[aps, prd, floatfix, twocolumn, superscriptaddress, nofootinbib]{revtex4-1}

\usepackage{latexsym}
\usepackage{amsmath}
\usepackage{amssymb}
\usepackage{amsfonts}

\usepackage{dsfont}
\usepackage{slashed}
\usepackage{bm}
\usepackage{urwchancal}

\usepackage[vcentermath]{youngtab}

\usepackage{color}
\definecolor{purple}{rgb}{0.5,0,0.5}
\definecolor{blue}{rgb}{0.0,0,0.9}
\definecolor{prdblue}{rgb}{0.133,0.118,0.498}

\usepackage{enumitem}
\usepackage{multirow}
\usepackage{subfigure}
\usepackage{textcomp}

\usepackage{supertabular} 
\usepackage{placeins}
\usepackage{epsfig}
\usepackage{graphicx}
\usepackage{soul} 

\begin{document}


\title{The ${\bf Z_b}$ structures in a constituent quark model coupled-channels calculation}

\author{Pablo G. Ortega}
\email[]{pgortega@usal.es}
\affiliation{Departamento de Física Fundamental and Instituto Universitario de F\'isica 
Fundamental y Matem\'aticas (IUFFyM), Universidad de Salamanca, E-37008 Salamanca, Spain}

\author{Jorge Segovia}
\email[]{jsegovia@upo.es}
\affiliation{Departamento de Sistemas F\'isicos, Qu\'imicos y Naturales, \\ Universidad Pablo de Olavide, E-41013 Sevilla, Spain}

\author{Francisco Fern\'andez}
\email[]{fdz@usal.es}
\affiliation{Grupo de F\'isica Nuclear and Instituto Universitario de F\'isica 
Fundamental y Matem\'aticas (IUFFyM), Universidad de Salamanca, E-37008 
Salamanca, Spain}

\date{\today}

\begin{abstract}
The $Z_b(10610)^\pm$ and $Z_b(10650)^\pm$ are two bottomonium-like structures discovered in the $\pi h_b(mP)$, $\pi \Upsilon(nS)$ and $B^\ast\bar B^{(\ast)}+h.c.$ invariant mass spectra, where $m=\{1,2\}$ and $n=\{1,2,3\}$. Their nature is puzzling due to their charge, which forces its minimal quark content to be $b\bar b u\bar d$ ($b\bar b d\bar u$). Thus, it is necessary to explore four-quark systems in order to understand their inner structure. Additionally, their strong coupling to channels such as $\pi \Upsilon$ and the closeness of their mass to $B^\ast\bar B^{(\ast)}$-thresholds stimulates a molecular interpretation. 
Within the framework of a constituent quark model which satisfactorily describes a wide range of properties of (non-)conventional hadrons containing heavy quarks, we perform a coupled-channels calculation of the $I^G(J^{PC})=1^+(1^{+-})$ hidden-bottom sector including $B^{(\ast)}\bar B^{\ast}+h.c.$, $\pi h_b$, $\pi \Upsilon$ and $\rho\eta_b$ channels.
We analyze the line shapes in the different channels, describing the $\Upsilon(5S)\to \pi B^{(*)}\bar B^{(*)}$ by means of the $^3P_0$ model. Since our description of the line shapes promising, we perform the same coupled-channels calculation for the $Z_b$'s with $J^{--}$, where $J=\{0,1,2\}$. This allows us to obtain a fair description of the corresponding line shapes.
The study of the analytic structure of the $S$-matrix suggests that the experimental $Z_b$ structures arise as a combination of several poles with $J^{PC}=0^{--}$, $1^{\pm-}$ and $2^{--}$ quantum numbers nearby the $B\bar B^\ast$ and $B^\ast\bar B^\ast$ thresholds.
\end{abstract}



\maketitle


\section{Introduction}
\label{sec:Introduction}

During almost two decades, tens of charmonium- and bottomonium-like states, the so-called XYZ mesons, have been observed at B-factories (BaBar, Belle and CLEO), $\tau$-charm facilities (CLEO-c and BESIII) and also proton-(anti)proton colliders (CDF, D0, LHCb, ATLAS and CMS). The common characteristic of most of these states is that, although their quantum  numbers are compatible with naive $Q\bar{Q}$ states ($Q$ either $c$- or $b$-quark), their  masses, static properties and decay patterns point out to more complex structures involving higher Fock-state components. 

It was not until $2011$ when undeniable evidences of exotic mesons with forbidden quantum numbers for a quark-antiquark pair were observed. The \emph{charged} bottomonium-like states $Z_b(10610)^\pm$ and $Z_b(10650)^\pm$ were identified by the Belle Collaboration~\cite{Belle:2011aa} as peaks in the invariant mass distribution of the $\pi^\pm h_b(mP)$ ($m=1,2$) and $\pi^\pm \Upsilon(nS)$ ($n=1,2,3$) subsystems when the $\Upsilon(10860)$ resonance decays into two pions plus an $h_b$ or $\Upsilon$. Shortly after, the Belle Collaboration~\cite{Adachi:2012cx, Garmash:2015rfd} confirmed their existence when investigating the elastic $B\bar{B}^{*}$ and $B^{*}\bar{B}^{*}$ channels. Moreover, the quantum numbers of the $Z_b$'s were analyzed to be $I^G(J^{PC})=1^+(1^{+-})$~\cite{Collaboration:2011gja} and so it appears in the Review of Particle Physics (RPP) of the Particle Data Group (PDG)~\cite{Zyla:2020zbs}. Note herein that a properly normalised $C$-odd combination of the $B\bar{B}^*$ and $\bar{B}B^*$ components must be understood from now on.

Since both the $Z_b(10610)^\pm$ and $Z_b(10650)^\pm$ contain a heavy $b\bar b$-pair, it is commonly accepted that the constraints from the heavy-quark flavor symmetry (HQFS) should be very accurate for these systems. In fact, one can predict their HQFS partners in the $c\bar c$ sector and assign them to the $Z_{c}(3900)^\pm/Z_c(3885)^\pm$ and $Z_{c}(4020)^\pm$ charmonium-like structures discovered in the $\pi J/\psi$, $\pi h_c$ and $D^\ast\bar D^{(\ast)}+h.c.$ invariant mass spectra~\cite{Ablikim:2013mio, Ablikim:2013wzq}. Their strong coupling to channels such as $\pi J/\psi$ and the closeness of their mass to $D^\ast\bar D^{(\ast)}$-thresholds stimulates a molecular interpretation and, among an extensive literature concentrated in these charmonium-like states, we performed in Ref.~\cite{Ortega:2018cnm} a coupled-channels calculation of the $I^G(J^{PC})=1^+(1^{+-})$ sector including $D^{(\ast)}\bar D^{\ast}+h.c.$, $\pi J/\psi$ and $\rho\eta_c$ channels in the framework of a constituent quark model. Therein, we observed that the meson-meson interactions are dominated by the non-diagonal $\pi J/\psi-D^\ast\bar D^{(\ast)}$ and $\rho\eta_c-D^\ast\bar D^{(\ast)}$ couplings indicating that the $Z_{c}(3900)^\pm/Z_c(3885)^\pm$ and $Z_{c}(4020)^\pm$ are unusual states. The study of the analytic structure of the $S$-matrix allowed us to conclude that the point-wise behavior of the line shapes in the $\pi J/\psi$ and $D\bar D^*$ invariant mass distributions is due to the presence of two virtual states that produce the $Z_c$ peaks. This calculation was recently extended to the open-strange sector with the description of the $Z_{cs}(3985)^-$~\cite{Ortega:2021enc}, with $c\bar c s\bar u$ minimal quark content.

Turning our attention back to the $Z_b$ states, both tetraquark~\cite{Ali:2011ug, Esposito:2014rxa, Maiani:2017kyi} and molecular~\cite{Bondar:2011ev, Cleven:2011gp, Nieves:2011zz, Zhang:2011jja, Yang:2011rp, Sun:2011uh, Ohkoda:2011vj, Li:2012wf, Ke:2012gm, Dias:2014pva} pictures are claimed to be in fair agreement with the experimental data. However, the $Z_b(10610)^\pm$ and $Z_b(10650)^\pm$ exotic states are located quite close to the $B\bar{B}^*$ and $B^*\bar{B}^*$ thresholds, respectively; and, moreover, these channels are the dominant decay channels of the $Z_b$'s. This provides a strong support for their molecular interpretation. Reference~\cite{Swanson:2014tra} argued that the $Z_b(10610)^\pm$ and $Z_b(10650)^\pm$ could be simple kinematical cusps; however, general arguments~\cite{Guo:2014iya} and explicit calculations~\cite{Hanhart:2015cua, Guo:2016bjq} demonstrate that narrow bumps as the $Z_b$'s signals in the $B\bar{B}^*$ and $B^*\bar{B}^*$ channels necessitate near-threshold poles.

One can imagine then that the literature related with the molecular scenario for the $Z_b$ states is already very rich, \emph{ i.e.} hadronic and radiative decays have been studied in Refs.~\cite{Voloshin:2011qa, Ohkoda:2012rj, Li:2012uc, Cleven:2013sq, Dong:2012hc, Li:2012as, Ohkoda:2013cea}, the contribution of the two $Z_b$ states to other scattering and production processes have been considered in Refs.~\cite{Chen:2011zv, Chen:2015jgl, Chen:2016mjn}, the heavy-quark spin partners of the $Z_b$'s have been discussed in Refs.~\cite{Mehen:2011yh, Ohkoda:2011vj, Valderrama:2012jv, HidalgoDuque:2012pq, Nieves:2012tt, Guo:2013sya, Karliner:2015ina, Baru:2017gwo}, the line shapes and poles position have been addressed in Refs.~\cite{Cleven:2011gp, Hanhart:2015cua, Guo:2016bjq, Kang:2016ezb, Nefediev:2017rfw}, predictions using different phenomenological models can be found in Refs.~\cite{Ding:2009zq, Li:2012wf}, and using QCD sum rules the $Z_b$'s were described in Refs.~\cite{Wang:2013daa, Wang:2014gwa, Agaev:2017lmc}.

Lattice QCD studies of the $Z_b$'s are scarce because the $I^G(J^{PC})=1^+(1^{+-})$ hidden-bottom sector presents severe challenges. If the L\"uscher method was used to study the poles of the Scattering matrix one should include, at least, seven coupled two-meson channels. Furthermore, the original L\"uscher approach for two-particle scattering is not valid above the three-particle threshold; though its extension to the three-body scattering processes is underway~\cite{Briceno:2019muc, Hansen:2020zhy}. Despite of these difficulties, preliminary lattice studies have been reported in Refs.~\cite{Peters:2016wjm, Peters:2017hon}. Moreover, a Born-Oppenheimer approximation~\cite{Braaten:2014qka, Brambilla:2017uyf}, inspired by the study of these systems in~\cite{Peters:2016wjm, Peters:2017hon}, has been recently applied to the $Z_b$ states within the framework of lattice QCD~\cite{Prelovsek:2019yae, Prelovsek:2019ywc}.

As a natural extension of our recent work on $Z_{c}(3900)^\pm/Z_c(3885)^\pm$ and $Z_{c}(4020)^\pm$~\cite{Ortega:2018cnm}, concluding that the point-wise behavior of the relevant line shapes is due to the presence of two virtual states in the $D^{(*)}\bar D^*$ channels with quantum numbers $J^{PC}=1^{+-}$, we analyze herein the $I^G(J^{PC})=1^+(0^{--})$, $1^+(1^{(\pm)-})$ and $1^+(2^{--})$ hidden-bottom sector in a coupled-channels scheme, including the following meson-meson thresholds: $B^{(\ast)}\bar B^{\ast}$, $\pi \Upsilon$, $\pi h_b$, and $\rho\eta_b$. The meson-meson interaction is described within the framework of a constituent quark model~\cite{Segovia:2013wma, Segovia:2016xqb} which has been successfully employed to explain the meson and baryon phenomenology from the light to the heavy quark sector~\cite{Vijande:2004he, Valcarce:2005rr, Valcarce:2008dr, Segovia:2008zza, Segovia:2009zz, Ortega:2011zza, Segovia:2011zza, Segovia:2015dia, Segovia:2016xqb, Yang:2019lsg}. Moreover, the $B^{(\ast)}\bar B^\ast$ interaction deduced from the model has been satisfactorily used to describe meson-meson~\cite{Ortega:2016mms, Ortega:2016pgg, Ortega:2017qmg} and meson-baryon~\cite{Ortega:2012cx, Ortega:2014eoa, Ortega:2014fha, Ortega:2016syt} molecular states such as the $X(3872)$ as a $D\bar D^\ast$ molecule coupled to $c\bar c(n^3P_1)$ states~\cite{Ortega:2009hj, Ortega:2012rs}.

The structure of the present manuscript is organized as follow. After this introduction, the theoretical framework is briefly discussed in Sec.~\ref{sec:Theory}. Our analysis of the obtained results: masses, decay widths and line shapes, can be found in Section~\ref{sec:Results}. Finally, we summarize and give some conclusions in Sec.~\ref{sec:Summary}.


\section{Theoretical formalism}
\label{sec:Theory}


\subsection{Constituent quark model}
\label{subsec:CQM}

In the absence of current quark masses, Quantum Chromodynamics (QCD) is invariant under chiral transformations. However, this symmetry is not realized in Nature indicating that it is spontaneously broken in QCD. Among its many consequences, the most outstanding are the development of a constituent quark mass which depends on the quark momentum, $M=M(q^2)$ with $M(q^2=\infty)=m_q$ being the current quark mass that appears in the QCD Lagrangian, and the existence of Goldstone-bosons whose exchanges between the light quarks produce interactions among them.

We use a so-called constituent quark model (CQM) which mimics the mentioned phenomena based on the following effective Lagrangian at low-energy~\cite{Diakonov:2002fq}:
\begin{equation}
{\mathcal L} = \bar{\psi}\, \Big[ i\, {\slash\!\!\! \partial} -M(q^{2})U^{\gamma_{5}} \Big]\,\psi  \,,
\label{eq:Lchiral}
\end{equation}
where $U^{\gamma_5} = e^{i\lambda _{a}\phi ^{a}\gamma _{5}/f_{\pi}}$ is the matrix of Goldstone-boson fields that can be expanded as follows
\begin{equation}
U^{\gamma _{5}} = 1 + \frac{i}{f_{\pi}} \gamma^{5} \lambda^{a} \pi^{a} - \frac{1}{2f_{\pi}^{2}} \pi^{a} \pi^{a} + \ldots \,,
\label{eq:MGB}
\end{equation}
in which the first term gives rise to the constituent quark mass, the second one describes the pseudoscalar-meson exchange interaction among quarks and the main contribution of the third term can be modeled by means of a scalar-meson exchange potential.

Another well-known non-perturbative phenomenon is the confinement of quarks inside hadrons. Lattice-QCD studies have demonstrated that multi-gluon exchanges produce an attractive linearly rising potential proportional to the distance between infinite-heavy quarks~\cite{Bali:2005fu}. However, the spontaneous creation of light-quark pairs from the QCD vacuum may give rise at the same scale to a breakup of the color flux-tube~\cite{Bali:2005fu}. We have tried to mimic these two phenomenological observations by the following expression:
\begin{equation}
V_{\rm CON}(\vec{r}\,)=\left[-a_{c}(1-e^{-\mu_{c}r})+\Delta \right]  (\vec{\lambda}_{q}^{c}\cdot\vec{\lambda}_{\bar{q}}^{c}) \,.
\label{eq:conf}
\end{equation}
Here, $a_{c}$ and $\mu_{c}$ are model parameters, and the SU(3) color Gell-Mann matrices are denoted as $\lambda^c$.

Finally, one expects that the dynamics of the bound-state system is governed by QCD perturbative effects at short inter-quark distances. This is taken into account by the one-gluon exchange potential derived from the following vertex Lagrangian
\begin{equation}
{\mathcal L}_{qqg} = i\sqrt{4\pi\alpha_{s}} \, \bar{\psi} \gamma_{\mu} 
G^{\mu}_c \lambda^c \psi \,,
\label{eq:Lqqg}
\end{equation}
with $\alpha_{s}$ an effective scale-dependent strong coupling constant which allows us to consistently describe light, strange and heavy quark sectors, and whose explicit expression can be found in, \emph{e.g.}, Ref.~\cite{Segovia:2008zz}.

A detailed physical background of the quark model can be found in Refs.~\cite{Segovia:2013wma, Segovia:2016xqb}. The model parameters and explicit expressions for the potentials can be also found therein. We want to highlight here that the interaction terms between light-light, light-heavy and heavy-heavy quarks are not the same in our formalism, \emph{ i.e.} while Goldstone-boson exchanges are considered when the two quarks are light, they do not appear in the other two configurations: light-heavy and heavy-heavy; however, the one-gluon exchange and confining potentials are flavor-blind.


\subsection{Resonating group method}
\label{subsec:RGM}

The Resonating Group Method (RGM)~\cite{Wheeler:1937zza} allows us to describe the strong interaction at the meson level. We are going to consider mesons as quark-antiquark clusters and the effective cluster-cluster interaction emerges from the aforementioned microscopic interaction among constituent quarks.

The wave function of a system composed of two mesons, $A$ and $B$, with distinguishable quarks, can be written as\footnote{Note that, for simplicity of the discussion presented herein, we have dropped off the spin-isospin wave function, the product of the two color singlets and the wave function that describes the center-of-mass 
motion.}
\begin{equation}
\langle \vec{p}_{A} \vec{p}_{B} \vec{P} \vec{P}_{\rm c.m.} | \psi 
\rangle = \phi_{A}(\vec{p}_{A}) \phi_{B}(\vec{p}_{B}) 
\chi_{\alpha}(\vec{P}) \,,
\label{eq:wf}
\end{equation}
where, \emph{e.g.}, $\phi_{A}(\vec{p}_{A})$ is the wave function of the meson $A$ with $\vec{p}_{A}$ the relative momentum between its quark and antiquark. The relative motion of the two mesons is taken into account by the wave function $\chi_\alpha(\vec{P})$.

The projected Schr\"odinger equation for the relative wave function can be written as follows:
\begin{align}
&
\left(\frac{\vec{P}^{\prime 2}}{2\mu}-E \right) \chi_\alpha(\vec{P}') + \sum_{\alpha'}\int d\vec{P}_{i}\, \Bigg[ {}^{\rm RGM}V_{D}^{\alpha\alpha'}(\vec{P}',\vec{P}_{i}) + \nonumber \\
&
+ {}^{\rm RGM}V_{R}^{\alpha\alpha'}(\vec{P}',\vec{P}_{i}) \Bigg] \chi_{\alpha'}(\vec{P}_{i}) = 0 \,,
\label{eq:Schrodinger}
\end{align}
where $E$ is the energy of the meson-meson system. The direct potential, ${}^{\rm RGM}V_{D}^{\alpha\alpha '}(\vec{P}',\vec{P}_{i})$, can be written as
\begin{align}
&
{}^{\rm RGM}V_{D}^{\alpha\alpha '}(\vec{P}',\vec{P}_{i}) = \sum_{i\in A, j\in B} \int d\vec{p}_{A'} d\vec{p}_{B'} d\vec{p}_{A} d\vec{p}_{B} \times \nonumber \\
&
\times \phi_{A}^{\ast}(\vec{p}_{A'}) \phi_{B}^{\ast}(\vec{p}_{B'}) 
V_{ij}^{\alpha\alpha '}(\vec{P}',\vec{P}_{i}) \phi_{A'}(\vec{p}_{A}) \phi_{B'}(\vec{p}_{B})  \,;
\end{align}
whereas ${}^{\rm RGM}V_{R}^{\alpha\alpha'}(\vec{P}',\vec{P}_{i})$ is the quark re-arrangement potential and represents a natural way to connect meson-meson channels with different quark content such as $\pi \Upsilon$ and $B\bar B^\ast$. It is given by
\begin{align}
{}^{\rm RGM}V_{R}^{\alpha\alpha'}(\vec{P}',&\vec{P}_{i}) = \int d\vec{p}_{A'}\, 
d\vec{p}_{B'}\, d\vec{p}_{A}\, d\vec{p}_{B}\, d\vec{P}\, \phi_{A}^{\ast}(\vec{p}_{A'}) \times \nonumber \\
&
\times  \phi_{B}^{\ast}(\vec{p}_{B'}) 
V_{ij}^{\alpha\alpha '}(\vec{P}',\vec{P}) P_{mn} \times \nonumber \\
&
\times \left[\phi_{A'}(\vec{p}_{A}) \phi_{B'}(\vec{p}_{B}) \delta^{(3)}(\vec{P}-\vec{P}_{i}) \right] \,,
\label{eq:Kernel}
\end{align}
where $P_{mn}$ is an operator that exchanges quarks between clusters. 

The Rayleigh-Ritz variational principle is used to determine the eigen-values and -functions of the two-body Schr\"odinger equation that describes mesons. This is because its simplicity and flexibility; however, the choice of basis to expand the wave function solution is of great importance. We use the Gaussian Expansion Method~\cite{Hiyama:2003cu} which provides enough accuracy and simplifies the subsequent evaluation of the needed matrix elements. With the aim of optimizing the Gaussian ranges employing a reduced number of free parameters, Gaussian trial functions, whose ranges are given by a geometrical progression, are introduced~\cite{Hiyama:2003cu}. This choice produces a dense distribution at short distances enabling better description of the dynamics mediated by short range potentials.

The coupled-channels RGM equation, Eq.~\eqref{eq:Schrodinger}, can be rewritten as a set of coupled Lippmann-Schwinger equations of the following form
\begin{align}
T_{\alpha}^{\alpha'}(E;p',p) &= V_{\alpha}^{\alpha'}(p',p) + \sum_{\alpha''} \int
dp''\, p^{\prime\prime2}\, V_{\alpha''}^{\alpha'}(p',p'') \nonumber \\
&
\times \frac{1}{E-{\cal E}_{\alpha''}(p^{''})}\, T_{\alpha}^{\alpha''}(E;p'',p) \,,
\end{align}
where $\alpha$ labels the set of quantum numbers needed to uniquely define a certain partial wave, $V_{\alpha}^{\alpha'}(p',p)$ is the projected potential that contains the direct and re-arrangement potentials, and ${\cal E}_{\alpha''}(p'')$ is the energy corresponding to a momentum $p''$, written in the nonrelativistic case as:
\begin{equation}
{\cal E}_{\alpha}(p) = \frac{p^2}{2\mu_{\alpha}} + \Delta M_{\alpha} \,.
\end{equation}
Herein, $\mu_{\alpha}$ is the reduced mass of the meson-meson system corresponding to the channel $\alpha$, and $\Delta M_{\alpha}$ is the difference between the meson-meson threshold and the one we shall take as a reference.

The coupled-channels Lippmann-Schwinger equation is solved using a generalization of the matrix-inversion method~\cite{Machleidt:1003bo} in order to include channels with different thresholds. Once the $T$-matrix is calculated, we determine the on-shell part which is directly related to the scattering matrix (in the case of non-relativistic kinematics):
\begin{equation}
S_{\alpha}^{\alpha'} = 1 - 2\pi i 
\sqrt{\mu_{\alpha}\mu_{\alpha'}k_{\alpha}k_{\alpha'}} \, 
T_{\alpha}^{\alpha'}(E+i0^{+};k_{\alpha'},k_{\alpha}) \,,
\end{equation}
with $k_{\alpha}$ the on-shell momentum for channel $\alpha$.

Since our aim is to explore within the same formalism the existence of states below and above meson-meson thresholds. All the potentials and kernels shall be analytically continue for complex momenta; this allows to find the poles of the $T$-matrix in any possible Riemann sheet. 

\subsection{Three-body production amplitude}\label{sec:3P0ampli}

The experimental line shapes analyzed in this manuscript are calculated in $e^+e^-$ collisions at the
center of mass energy of the $\Upsilon(10860)$ meson. Thus, it is reasonable to think
that a large portion of the $Z_b$ candidates emerges from mechanisms in which the $\Upsilon(10860)$ dessintegrates into a 
pion plus a $Z_b$.

In order to shed some light into the $Z_b$ decay line shapes and to analyze possible contribution from
alternative $J^{PC}$ sectors, it is relevant to properly account for
the momentum dependence of the $Z_b's$ production vertex, as it would be questionable to assume a constant
production vertex for channels in a relative $P$ wave, as the $\pi h_b(mP)$ channels in the $J^{PC}=1^{+-}$ sector.

\begin{figure}[!t]
\centering
\includegraphics[width=0.35\textwidth]{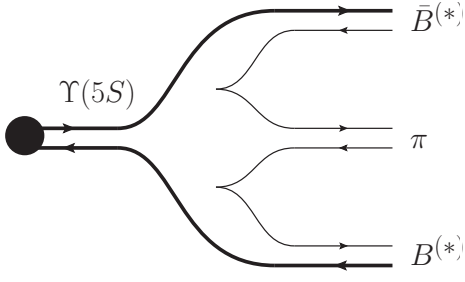}
\caption{\label{fig:3P0} Diagram for $\Upsilon(5S)\to \pi(B^{(*)}\bar B^{(*)}$ in the $^3P_0$ formalism.}
\end{figure}

In a molecular interpretation for the $Z_b$ candidates, the reaction $\Upsilon \to \pi(M_1M_2)$ involves the creation of two quark-antiquark
pairs (see Fig.~\ref{fig:3P0}), where $M_1M_2$ is the final two-body state of the coupled-channels calculation, {\it i.e.},
$M_1M_2=\{\pi\Upsilon,\pi h_b,B\bar B^*,B^*\bar B^*\}$. 
A simple approach for such reaction can be obtained from the quark-antiquark pair creation $^3P_0$ 
model~\cite{Micu:1968mk,LeYaouanc:1972vsx,Weng:2018ebv}. The transition operator for the aforementioned reaction 
can be written as,

\begin{align}
T &=9\gamma^{\prime\,2}\sum_\mu \int d^3 p_{\mu_1} d^3p_{\nu_1}d^3 p_{\mu_2} d^3p_{\nu_2} 
\,\delta^{(3)}(p_{\mu_1}+p_{\nu_1})\nonumber \\&\times
\,\delta^{(3)}(p_{\mu_2}+p_{\nu_2})
\left[ \mathcal Y_1\left(\frac{p_{\mu_1}-p_{\nu_1}}{2}\right) b_{\mu_1}^\dagger(p_{\mu_1})
d_{\nu_1}^\dagger(p_{\nu_1}) \right]_{\{1,0,1,0\}}  \nonumber \\&\times
\left[ \mathcal Y_1\left(\frac{p_{\mu_2}-p_{\nu_2}}{2}\right) b_{\mu_2}^\dagger(p_{\mu_2})
d_{\nu_2}^\dagger(p_{\nu_2}) \right]_{\{1,0,1,0\}}
\label{TBon}
\end{align}
where $\mu_i$ ($\nu_i=\bar \mu_i$) are the quark (antiquark) quantum numbers for the pair $i=\{1,2\}$, and
$\gamma'=2^{5/2}\sqrt{2} \pi^{1/2}\gamma$ with $\gamma= \frac{g}{2m}$ is a dimensionless constant 
that gives the strength of 
the $q\bar q$ pair creation from the vacuum. The subscript brakets in the operator represent the 
$\{C,I,S,J\}$ quantum numbers of the created $q\bar q$ pair.

From this transition operator we can obtain the potential $V_{^3P_0}(k)$, 
where $k$ is the relative momentum of the final two mesons in the coupled-channels calculation. 
For this potential, we employ the same meson wave functions obtained from the CQM model, as done for the coupled-channels calculation. A relevant key of this approach for the $Z_b$ production is that the $^3P_0$ mechanism only 
allows the $\Upsilon(10860)\to\pi (B^{(*)}\bar B^{(*)})$ transitions, so the production of hidden-bottom 
channels such as $\pi\Upsilon(nS)$ or $\pi h_b(mP)$ only arise  through a further quark rearrangement process, {\it i.e.}, $B^{(*)}\bar B^{(*)}\to \pi\Upsilon(nS)$ or $B^{(*)}\bar B^{(*)}\to \pi h_b(mP)$.

\subsection{Production line shapes}\label{sec:lineshape}

\begin{figure}[!t]
\centering
\includegraphics[width=0.48\textwidth]{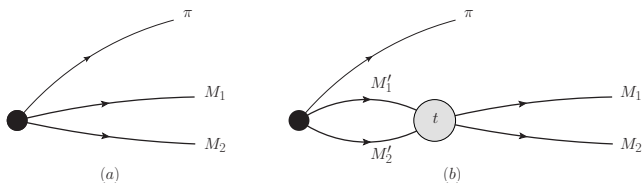}
\caption{\label{fig:Prod} Background (a) and rescattering (b) contributions for the production of the two-meson $M_1M_2$ channel, ${\cal M}$, as denoted in Eq.~\eqref{eq:amplitude}. Hidden-botom channels are only produced through the rescattering diagram (b).}
\end{figure}

The $Z_b\to M_1M_2$ production is described diagramatically as in Fig.~\ref{fig:Prod}, and it can be written in terms of the following Lorentz-invariant production amplitude:
\begin{align}
\mathcal{M^\beta}(m_{M_1M_2}) &= {\cal A}^\beta e^{i\,\theta_\beta} V_{^3P_0}(k^\beta)\nonumber \\
&
-\sum_{\beta'}{\cal A}^{\beta'} e^{i\,\theta_{\beta'}}\int d^3p\frac{V_{^3P_0}(p) t^{\beta'\beta}(p,k^\beta,E)}{p^2/2\mu-E-i0} \,.
\label{eq:amplitude}
\end{align}
where $\beta^{(\prime)}$ denotes a given $M_1M_2$ channel in the coupled-channels calculation and the amplitude is summed up over the different $J^{PC}$ considered in this work. 
In order to explore the description of line shapes, we add the amplitudes ${\cal A}^\beta$ and phases $\theta_\beta$ to take into account different production weights for each channel $M_1M_2$. It is worth reminding that the background contribution only affects the $B\bar B^*$ and $B^*\bar B^*$ channels, as the hidden-bottom channels are OZI-suppressed.

A resonance in its rest frame, which decays into a three-body final state $\pi(M_1M_2)$, has a differential decay rate given by
\begin{equation}
d\Gamma = \frac{(2\pi)^4}{2M} \, |{\cal M}|^2 \, d\Phi(P;p_\pi,p_{M_1},p_{M_2}) \,,
\end{equation}
where ${\cal M}$ is the Lorentz-invariant production amplitude from the $^3P_0$ amplitude described above, defined in Eq.~\eqref{eq:amplitude}, and $d\Phi$ is the three-body phase-space which can be written as
\begin{equation}
d\Phi(P;p_1,p_2,p_3)=\delta^{(4)}\left(P-\sum_i p_i \right) \prod_i \frac{d^3p_i}{(2\pi)^3 2 E_i} \,.
\end{equation}

Therefore, we can express $d\Gamma$ in terms of the invariant mass spectrum of the two-meson channel, $m_{M_1M_2}$, as
\begin{equation}
d\Gamma = \frac{1}{(2\pi)^5}\frac{k_{M_1M_2}\, k_{\pi Z_b}}{16\,s} \, |{\cal M}|^2 \, dm_{M_1M_2}\, d\Omega_{M_1M_2}\, d\Omega_{\pi Z_b} \,,
\end{equation}
where $(k_{M_1M_2},\Omega_{M_1M_2})$ is the on-shell momentum of the $M_1M_2$ pair and $\Omega_{\pi Z_b}$ is the solid angle of the $\pi$ in the center-of-mass reference frame of $\pi(M_1M_2)$ at energy $\sqrt{s}$. 
The on-shell momenta are given by
\begin{subequations}
\begin{align}
k_{\pi Z_b} &= \frac{\lambda^{1/2}(\sqrt{s},m_{M_1M_2},m_\pi)}{2\sqrt{s}} \,, \label{ec:kmomZpi} \\
k_{M_1M_2} &= \frac{\lambda^{1/2}(m_{M_1M_2},m_A,m_B)}{2m_{M_1M_2}} \,, \label{ec:kmomDD}
\end{align}
\end{subequations}
where $\lambda(M,m_1,m_2)=[(M^2-m_+^2)(M^2-m_-^2)]$, with $m_\pm=m_1\pm m_2$. Integrating the angles, we have
\begin{equation}\label{eq:decay3P0}
d\Gamma = \frac{1}{(2\pi)^3}\, \frac{k_{M_1M_2} k_{\pi Z_c}}{4\,s}\, |\overline{\cal M}^\beta(m_{M_1M_2})|^2 \, dm_{M_1M_2} \,,
\end{equation}
with $\beta$ the quantum numbers of the channel $M_1M_2$ and $\overline{\cal M}^\beta$ is averaged over $\Upsilon(5S)$ spin states.

Experimentalists actually measure events in the $\sigma(e^+e^-\to \pi^\pm Z_b^\mp)\times \mathcal{B}(Z_b^\mp \to M_1M_2)$ process; to describe the data, we need to add a normalization factor to translate the decay rate into events:
\begin{equation}
N(m_{M_1M_2}) = \mathcal{N}_{M_1M_2}\times \frac{d\Gamma_{Z_b\to M_1M_2}}{dm_{M_1M_2}} \,.
\end{equation}

In order to fit $\{{\cal A}_{M_1M_2},\theta_{M_1M_2},\mathcal{N}_{M_1M_2}\}$ for each channel, we minimize a global $\chi^2$ function using all the available experimental data for channels $\pi \Upsilon$, $\pi h_b$, $B\bar B^\ast$ and $B^\ast\bar B^\ast$:
\begin{equation}
\label{eq:chisquare}
\chi^2(\{{\cal A,\theta,N}\}) = \sum_i \bigg(\frac{N^{\rm the}(x_i)-N^{\rm exp}(x_i)}{\sigma_i^{\rm exp}}\bigg)^2 \,.
\end{equation}

The theoretical lineshapes are, then, convolved with a $\sigma=6$ MeV Gaussian, to take into account the experimental resolution~\cite{Belle:2011aa,Adachi:2012cx,Garmash:2015rfd}.

\section{Results}
\label{sec:Results}

It is worth emphasizing firstly that Heavy Flavour Symmetry (HFS) dictates the $B^{(\ast)}\bar B^\ast$ and $D^{(\ast)}\bar D^\ast$ interactions are practically identical. This happens because the $B^{(\ast)}$ and $D^{(\ast)}$ wave functions are similar and, regarding the meson-meson interaction, the heavy quark $c$, or $b$, acts as an spectator and the dynamics is govern by the light quark pair. However, the available kinetic energy of the involved channels is reduced in the bottom sector due to the larger mass of the $b$-quark, favoring the creation of bound states.

We begin performing a coupled-channels calculation for the $I^G(J^{PC})=1^+(1^{+-})$ sector, including the following channels:  $\pi\Upsilon(1S)$ ($9597.6$ MeV/$c^2$), $\pi h_b(1P)$ ($10036.6$ MeV/$c^2$), $\pi\Upsilon(2S)$ ($10160.5$ MeV/$c^2$), $\rho\eta_b$ ($10174.3$ MeV/$c^2$), $\pi h_b(2P)$ ($10397.1$ MeV/$c^2$), $\pi\Upsilon(3S)$ ($10492.5$ MeV/$c^2$), $B\bar B^\ast$ ($10604.1$ MeV/$c^2$) and $B^\ast\bar B^\ast$ ($10649.3$ MeV/$c^2$). Note that, in our model, the $\pi h_b$ channel couples only, and very weakly, to the $B\bar B^{\ast}$ one when the quantum numbers are $I^G(J^{PC})=1^+(1^{+-})$. In fact, the binding-energy contribution of this channel would come from the $1/m_q m_{\bar q}$ suppressed spin-orbit term, which would be able to flip the spin of the heavy-quark pair.

The first column of Tables~\ref{tab:Zbpoles} and~\ref{tab:Zbpoles2} show our findings for the $I^G(J^{PC})=1^+(1^{+-})$ hidden-bottom sector. Contrary to the $Z_c$ case~\cite{Ortega:2018cnm}, the $B^{(\ast)}\bar B^\ast$ interaction is strong enough to produce two real bound states below the $B\bar B^\ast$ and the $B^\ast\bar B^\ast$ thresholds. Tables~\ref{tab:Zbpoles} and~\ref{tab:Zbpoles2} collect the largest meson-meson component in the wave function of our candidates for the $Z_b(10610)^\pm$ and $Z_b(10650)^\pm$ states; one can see that the dominant wave-function's component is $B\bar B^\ast$ and $B^\ast\bar B^\ast$, respectively, which facilitates their interpretation as $B^{(\ast)}\bar B^\ast$ molecular states. 

Our prediction for the $Z_b$'s total decay widths, broken down into the different meson-meson final state contributions, are also presented in Tables~\ref{tab:Zbpoles} and~\ref{tab:Zbpoles2}. We are not able to reproduce the total decay widths reported by the Belle collaboration for neither $Z_b(10610)^\pm$ nor $Z_b(10650)^\pm$. Moreover, their coupling to the $\pi h_b$ channel is very little (not collected in the Table) with the $Z_b$'s interpreted as molecular states of $B^\ast\bar{B}^\ast$ with quantum numbers $I^G(J^{PC})=1^+(1^{+-})$.

\begin{table}[!t]
\begin{tabular}{lrr|rrrrr}
\hline
\hline
\multicolumn{2}{c}{$J^{PC}$}  & & $1^{+-}$ &  $0^{--}$  & $1^{--}$  & $2^{--}$ & \\
\hline
Mass  & (MeV) & & $10602.5$ & $10603.7$ & $10600.2$ & $10599.0$ & \\ 
Width & (MeV) & & $0.11$ & $6.23$ & $41.32$ & $22.06$ &\\
${\cal P}_{B\bar B^*}$ & (\%) & & $96.20$ & $99.98$ & $99.99$ & $87.21$ &\\
\hline
$\Gamma_{B\bar B}$ & (MeV) & & $0$ & $0$ & $0$ & $0$ &\\
$\Gamma_{B\bar B^*}$ & (MeV) & & $0$ & $0$ & $0$ & $0$ &\\
$\Gamma_{\pi\Upsilon(1S)}$ & (MeV) & & $0.10$ & $0.69$ & $10.07$ & $5.43$ & \\
$\Gamma_{\pi\Upsilon(2S)}$ & (MeV) & & $0.01$ & $1.55$ & $26.46$ & $13.06$ &\\
$\Gamma_{\pi\Upsilon(3S)}$ & (MeV) & & $0$ & $0.11$ & $2.37$ & $1.06$ &\\
$\Gamma_{\rho \eta_b}$ & (MeV) & & $0.005$ & $3.88$ & $2.42$ & $2.51$ & \\
$\Gamma_{\pi h_b(1P)}$ & (MeV) & & $0$ & $0$ & $0$ & $0$ & \\
$\Gamma_{\pi h_b(2P)}$ & (MeV) & & $0$ & $0$ & $0$ & $0$ & \\
\hline
\hline
\end{tabular}
\caption{\label{tab:Zbpoles} The $Z_b$ states parameters from the $S$-matrix poles nearby to the $B\bar B^*$ threshold for different $J^{PC}$ sectors.
The $1^{+-}$ state is a bound state, while the rest are virtual states.}
\end{table}

\begin{table}[!t]
\begin{tabular}{lrr|rrrr}
\hline
\hline
\multicolumn{2}{c}{$J^{PC}$}  & & $1^{+-}$  & $1^{--}$  & $2^{--}$ & \\
\hline
Mass  & (MeV) & & $10649.4$ & $10651.1$ & $10647.1$  &\\ 
Width & (MeV) & & $0.76$ & $20.04$ & $7.63$ & \\
${\cal P}_{\rm B^*\bar B^*}$ & (\%) & & $67.99$ & $87.51$ & $94.29$ & \\
\hline
$\Gamma_{B\bar B}$ & (MeV) & & $0$ & $1.78$ & $0$ &\\
$\Gamma_{B\bar B^*}$ & (MeV) & & $0.32$ & $0$ & $7.63$  &\\
$\Gamma_{B^*\bar B^*}$ & (MeV) & & $0$ & $16.45$ & $0$ &\\
$\Gamma_{\pi\Upsilon(1S)}$ & (MeV) & & $0.40$ & $0$ & $0.001$ & \\
$\Gamma_{\pi\Upsilon(2S)}$ & (MeV) & & $0.04$ & $0$ & $0.001$ & \\
$\Gamma_{\pi\Upsilon(3S)}$ & (MeV) & & $0$ & $0$ & $0$ & \\
$\Gamma_{\rho \eta_b}$ & (MeV) & & $0.003$ & $0$ & $0$ & \\
$\Gamma_{\pi h_b(1P)}$ & (MeV) & & $0$ & $1.44$ & $0$  &\\
$\Gamma_{\pi h_b(2P)}$ & (MeV) & & $0$ & $0.37$ & $0$  &\\
\hline
\hline
\end{tabular}
\caption{\label{tab:Zbpoles2} The $Z_b^\prime$ states parameters from the $S$-matrix poles nearby the $B^*\bar B^*$ threshold for different $J^{PC}$ sectors.
The $1^{+-}$ is a bound state, the $1^{--}$ is a resonance and the $2^{--}$ is a virtual state.
}
\end{table}

With this theoretical description of the $Z_b$'s at hand, we computed their production rates and compare them with the experimental data. We consider that the $Z_b(10610)^\pm$ and $Z_b(10650)^\pm$ states are coming from a $\Upsilon(5S)\to\pi(B^{(*)}\bar B^{(*)})\to \pi (M_1M_2)$ vertex described by means of a $^3P_0$ amplitude with a center-of-mass energy $\sqrt{s}=10.865$ GeV, according to Refs.~\cite{Belle:2011aa, Bondar:2013nea, Adachi:2012cx, Garmash:2015rfd}.

On a first step, we can evaluate the goodness of the proposed $^3P_0$ production amplitude by analyzing the decay branching ratios of $\Upsilon(5S)\to \pi B^{(*)}\bar B^{(*)}$, integrating Eq.~\eqref{eq:decay3P0}. Experimental values for such branchings have been measured in Ref.~\cite{Drutskoy:2010an}:

\begin{align}
 &{\cal B}(\Upsilon(10860)\to \pi B\bar B) = (0.0\pm 1.2\pm 0.3)\%,\nonumber\\
 &{\cal B}(\Upsilon(10860)\to \pi (B\bar B^*+B^*\bar B)) = (7.3^{+2.3}_{-2.1}\pm0.8)\%,\nonumber\\
 &{\cal B}(\Upsilon(10860)\to \pi B^*\bar B^*) = (1.0^{+1.4}_{-1.3}\pm 0.4)\%.
\end{align}
over a total width of $\Gamma_{\Upsilon}^{\rm exp}=37\pm 4$ MeV~\cite{Zyla:2020zbs}.

The only parameter of the production amplitude is the value of $\gamma$ of the $^3P_0$ model. 
We will take the same value fixed for two-body decays of bottomonium studied in Ref.~\cite{Segovia:2016xqb}, 
considering a $10\%$ uncertainty, {\it viz.} $\gamma=0.205\pm0.020$. 
Without considering rescattering contributions in Eq.~\eqref{eq:decay3P0}, such value of $\gamma$ gives,

\begin{align}
 &{\cal B}(\Upsilon(10860)\to \pi B\bar B) = 2 \pm 1\%,\nonumber\\
 &{\cal B}(\Upsilon(10860)\to \pi (B\bar B^*+B^*\bar B)) = 6^{+3}_{-2}\%,\nonumber\\
 &{\cal B}(\Upsilon(10860)\to \pi B^*\bar B^*) = 3 \pm 1\%.
\end{align}
which are in good agreement with the experimental values. If we analyze the contribution of each $B^{(*)}\bar B^{(*)}$ partial wave to the total (Table~\ref{tab:decays}), we see that the $S$ and $P$ waves are of the same order, so we expect a similar contribution of $S$ and $P$ waves in the line shapes. 

\begin{table}[!t]
\begin{tabular}{c|ccc}
\hline\hline
 $J^{PC}$  & Channel & $^{2S+1}L_J$ & Width [MeV] \\ \hline
 $0^{--}$ & $B\bar B^*$ & $^3P_0$ & $0.51$ \\
 $1^{--}$ & $B\bar B$  & $^1P_1$ & $0.75$ \\
          & $B\bar B^*$ & $^3P_1$ & $0.39$ \\
          & $B^*\bar B^*$ & $^1P_1$ & $0.05$ \\
          & $B^*\bar B^*$ & $^5P_1$ & $0.24$ \\
$1^{+-}$ & $B\bar B^*$ & $^3S_1$ & $0.61$ \\
          & $B^*\bar B^*$ & $^3S_1$ & $0.21$ \\
$2^{--}$ & $B\bar B^*$ & $^3P_2$ & $0.59$ \\
          & $B^*\bar B^*$ & $^5P_2$ & $0.71$ \\ \hline\hline
\end{tabular}
\caption{\label{tab:decays} $\Upsilon(10860)\to \pi B^{(*)}\bar B^{(*)}$ decays within the $^3P_0$ model for different partial waves of the $B^{(*)}\bar B^{(*)}$ pair, using $\gamma=0.205$. }
\end{table}

Hence, as the production through a $P$ wave is significant, in order to deepen our understanding of the $Z_b$'s inner structure, we have performed three further coupled-channels computations for the quantum numbers $J^{--}$, with $J=\{0,1,2\}$.  Indeed, this allows us to include the $\pi h_b(mP)$ $(m=1,\,2)$ channels in our calculation in a relative $S$ wave in the $1^{--}$ sector~\footnote{Note, however, that in this sector the $\pi h_b$ channel does couple to the $B^*\bar B^*$ $^1P_1$ and $^5P_1$ partial waves through tensor potentials, but they are decoupled from the $B\bar B^*$ structure.}. For completeness, we add the $B\bar B$ channel in the coupled-channels calculation for those sectors.

\begin{figure}[!t]
\centering
\includegraphics[width=0.48\textwidth]{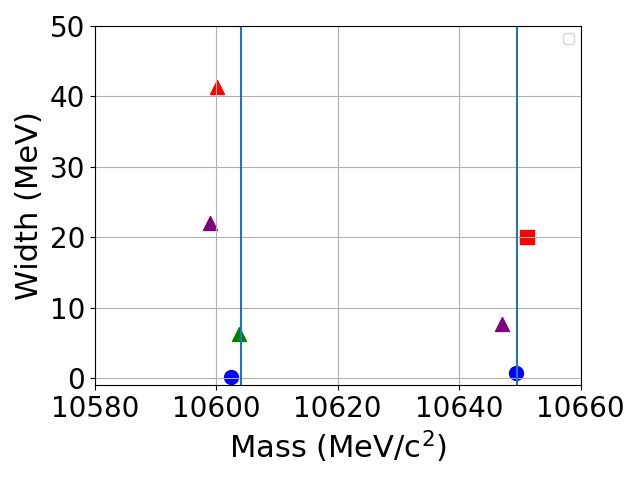}
\caption{\label{fig:poles} Position of $S$-matrix poles described in this work, as detailed in Tabs.~\ref{tab:Zbpoles} and~\ref{tab:Zbpoles2}. Circles denote bound states, triangles are virtuals and squares are resonances. Colors indicate the $J^{PC}$ sector: $1^{+-}$ in blue, $1^{--}$ in red, $0^{--}$ in green and $2^{--}$ in purple. Vertical lines represent the $B\bar B^*$ and $B^*\bar B^*$ thresholds.}
\end{figure}

One can see in Tables~\ref{tab:Zbpoles} and~\ref{tab:Zbpoles2} the $S$-matrix pole structure for the $J^{--}$ hidden-bottom sector, also summarized in Fig.~\ref{fig:poles}. As the $B^{(\ast)}B^\ast$ can only be in a relative $P$-wave for the $J^{--}$, we mostly obtain virtual states below the $B^*\bar B^{(*)}$ thresholds. A near-threshold resonance above $B^*\bar B^*$ channel is found for the $1^{--}$ sector, with mass close to the $Z_b(10650)^\pm$ state. All the states predicted have a dominant $B^{(\ast)}B^\ast$ component in their wave functions and thus their interpretation as molecular states is presumably. 

The computed decay widths of the three states, and their decomposition into the different meson-meson channels, are also collected in Tables~\ref{tab:Zbpoles} and~\ref{tab:Zbpoles2}. We are predicting the same order of magnitude, slightly higher, for the total decay widths reported experimentally: $(18.4\pm2.4)\,\text{MeV}$ for the $Z_b(10610)^\pm$, and $(11.5\pm2.2)\,\text{MeV}$ for the $Z_b(10650)^\pm$~\cite{Zyla:2020zbs}. Moreover, the most important decay channels are $\pi\Upsilon(nS)$ with $n=1,\,2,\,3$ for the $Z_b(10610)^\pm$ assignments, and $B^\ast \bar B^\ast$ for the three candidates of the $Z_b(10650)^\pm$. Let us finally mention here that the distinct states around the $B^\ast \bar B^\ast$ and $B \bar B^\ast$ thresholds with different quantum numbers $J^{PC}$ could be being interpreted as single $Z_b(10610)^\pm$ and $Z_b(10650)^\pm$ resonances in the experiments.

\begin{figure}[!t]
\centering
\includegraphics[width=0.48\textwidth]{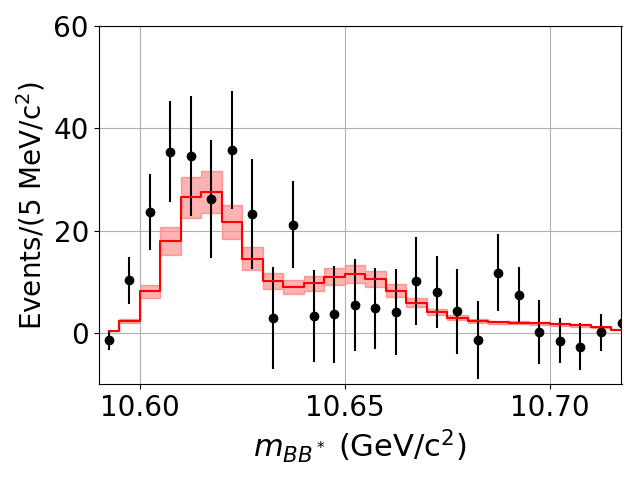}
\includegraphics[width=0.48\textwidth]{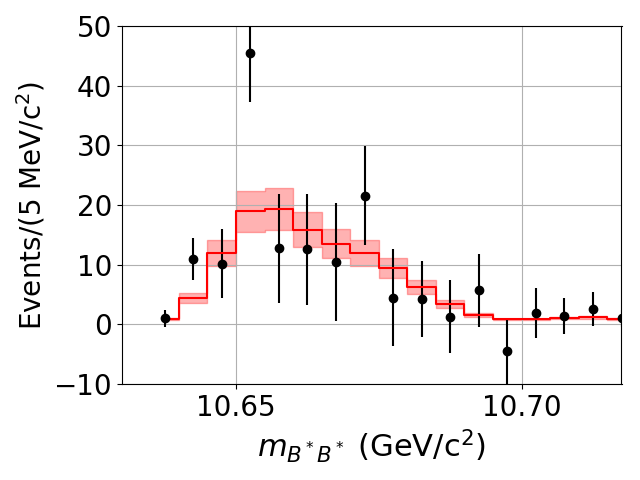}
\caption{\label{fig:ProdBB} Line-shape of the $B\bar B^*$ (upper pannel) and $B^*\bar B^*$ (lower pannel) channels. The red line shows the theoretical prediction. The line is convoluted  considering the experimental resolution of $6$ MeV. The black solid points are the background-sustracted experimental data from Ref.~\cite{Garmash:2015rfd}, whose statistical uncertainty is represented by a vertical line.}
\end{figure}

\begin{figure*}[!t]
\includegraphics[width=.45\textwidth]{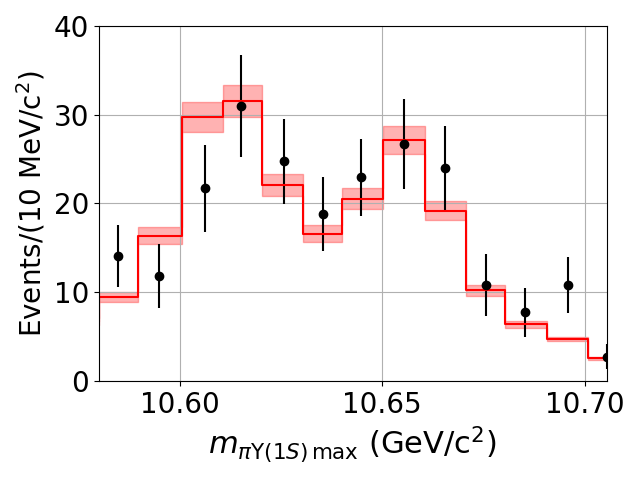}
\includegraphics[width=.45\textwidth]{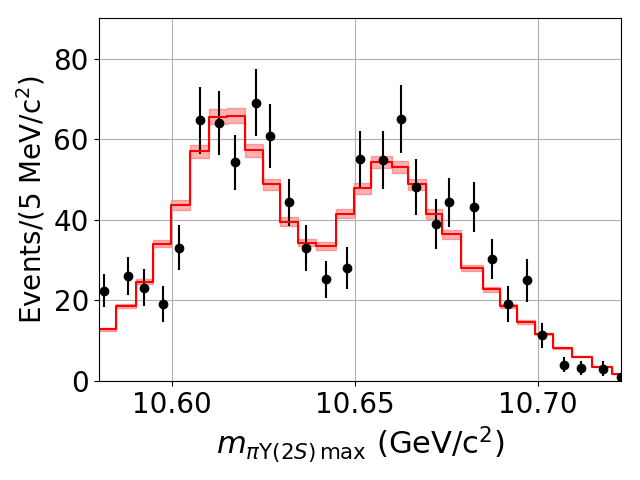}
\includegraphics[width=.45\textwidth]{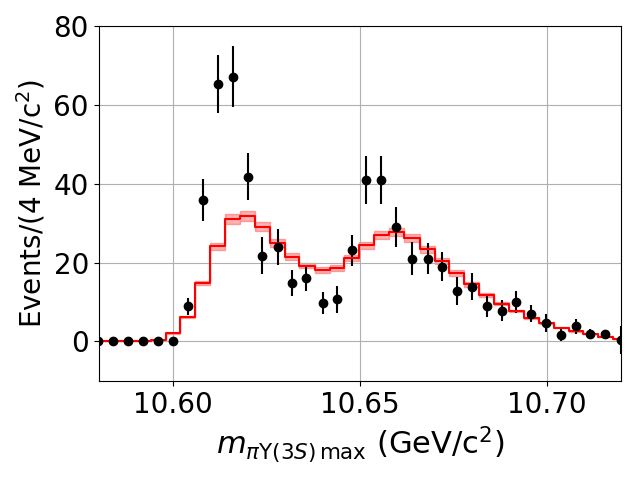}
\includegraphics[width=.45\textwidth]{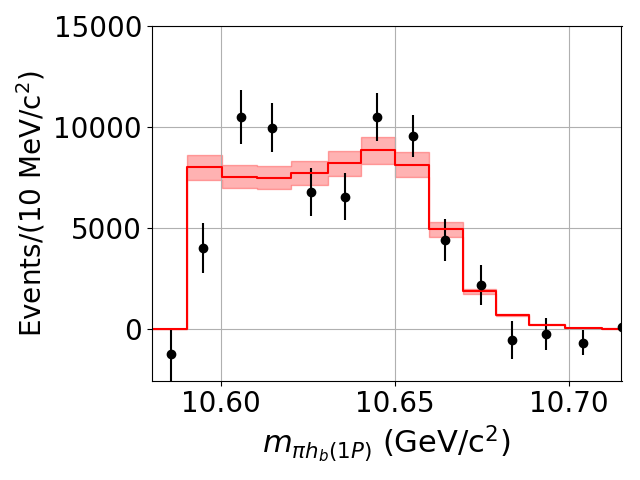}
\includegraphics[width=.45\textwidth]{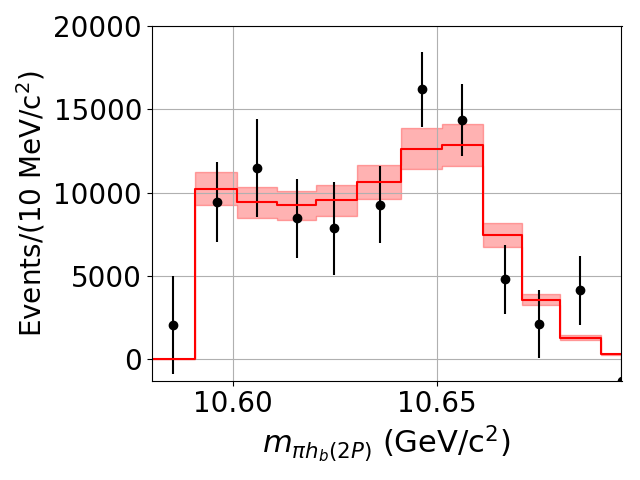}
\caption{\label{fig:line3} Line shapes of the $\pi\Upsilon(nS)$ $(n=1,\,2,\,3)$ and $\pi h_b(mP)$ $(m=1,\,2)$ channels (see legend in the abscissa axis). Same legend as Fig.~\ref{fig:ProdBB}. The black solid points are the experimental data from Refs.~\cite{Belle:2011aa, Bondar:2013nea, Adachi:2012cx}, whose statistical uncertainty is represented by a vertical line.}
\end{figure*}

The point-wise behavior of the line shapes for an invariant mass between $10.55$ and $10.75$ GeV can be seen in Fig.~\ref{fig:ProdBB} for the $B\bar B^\ast$ and $B^\ast\bar B^\ast$ final decay channels, and in Fig.~\ref{fig:line3} for the $\pi\Upsilon(nS)$ $(n=1,\,2,\,3)$ and $\pi h_b(mP)$ $(m=1,\,2)$ final decay channels. 

We find a good agreement with the $B^{(*)}\bar B^*$ experimental data in Fig.~\ref{fig:ProdBB}, due to the production from all sectors considered in this work. The $J^{--}$ poles close to the experimental $Z_b(10610)^\pm$ and $Z_b(10650)^\pm$ thresholds influence the $B\bar B^*$ and $B^*\bar B^*$ line shapes, respectively. The $Z_b(10650)^\pm$ peak in the $B\bar B^*$ line shape is seen as a small bump around $10.65$ GeV, smeared by the detector resolution.

The description of the line shapes for hidden-bottom channels $\pi\Upsilon(nS)$ and $\pi h_b(mP)$ is shown in Fig.~\ref{fig:line3}. 
It is worth remarking that all these channels are produced solely through rearrangement potentials from the $B^{(*)}\bar B^{(*)}$ channels. The $\pi\Upsilon(nS)$ ($n=1,2,3$) channels show two bumps which can be assigned to the two $Z_b$'s structures. A general good agreement is found for these line shapes, though the intensity of the peaks for the  $\pi\Upsilon(3S)$ channel is somehow lower than the experimental data.

Regarding the $\pi h_b(1P)$ and $\pi h_b(2P)$ line shapes, Figure~\ref{fig:line3}, the agreement to the experimental data is fair, showing a dominance of the second peak, which mostly originates from the  $1^{--}$ sector. The line shape shows a nearly constant production  with the invariant mass until the $B^\ast B^\ast$ mass threshold and then it falls-off quickly to zero, describing the experimental $Z_b(10650)^\pm$ structure. The first peak, corresponding to the $Z_b(10610)^\pm$ structure is, however, smeared, which points to a weaker $B\bar B^*\to\pi h_b(mP)$ interaction in the $1^{+-}$ sector compared to the $1^{--}$  $B^*\bar B^*\to\pi h_b(mP)$.


\section{Summary}
\label{sec:Summary}

The $Z_c$'s are very peculiar states, different from other potential molecular candidates of the hidden-charm spectrum. In order to clarify their inner structure, within the framework of a constituent quark model that satisfactorily describes a wide range of properties of (non-)conventional hadrons containing heavy quarks, we performed in Ref.~\cite{Ortega:2018cnm} a coupled-channels calculation of the $I^G(J^{PC})=1^+(1^{+-})$ $c\bar c u\bar{d}$ ($c\bar c d\bar{u}$) sector around the mass energies of the $Z_c(3900)^\pm/Z_c(3885)^\pm$ and $Z_c(4020)$ signals. We found that the diagonal interaction between the $D^{(*)}\bar D^{(*)}$ channels is, too, suppressed to develop resonances, being the non-diagonal rearrangement interaction due to the coupling with other meson-meson channels responsible for their virtual-state nature, needed to describe the relevant line shapes. The same approach has also been applied to the $c\bar c s\bar u$ sector pursuing a description of the $Z_{cs}(3985)^-$~\cite{Ortega:2021enc}.

We have naturally extended herein our analysis to the the $Z_b(10610)^\pm$ and $Z_b(10650)^\pm$ signals observed in 2011 by the Belle Collaboration, \emph{viz.} a coupled-channels calculation of the $I^G(J^{PC})=1^+(1^{+-})$ $b\bar b u\bar{d}$ ($b\bar b d\bar{u}$) sector, including $B^{(\ast)}\bar B^{\ast}$, $\pi \Upsilon(nS)$ ($n=1,2,3$), $\pi h_b(mP)$ ($m=1,2$) and $\rho \eta_b$ channels, has been performed within the same constituent quark model framework. In this initial calculation, we found that the $\pi h_b$ channel coupling to $B\bar B^*$ was rather small (due to the supressed spin-orbit term), and zero for the $B^*\bar B^*$ channel when only the quantum numbers $I^G(J^{PC})=1^+(1^{+-})$ are available for the $Z_b(10610)^\pm$ and $Z_b(10650)^\pm$ states. Thus, we perform the same coupled-channels calculation but considering too the following quantum numbers: $J^{PC}=0^{--}$, $1^{--}$ and $2^{--}$ for the $Z_b$'s. This preference is based on two main reasons: On the one hand, it allows us to include the $\pi h_b$ final states in $S$-wave, with a non-zero coupling to the $B^*\bar B^*$ channel. On the other hand, the description of the $\Upsilon(10860)\to \pi^\pm Z_b^\mp$ vertex using the $^3P_0$ model points to a significant contribution from $P$-wave $B^{(*)}\bar B^{(*)}$ channels. 

Several $S$-matrix poles were identified around the $B\bar B^*$ and $B^*\bar B^*$ thresholds, with masses and widths compatible with the  $Z_b(10610)^\pm$ and $Z_b(10650)^\pm$ resonances, and we obtained a fair description of the corresponding line shapes due to the combination of the production amplitudes of all those quantum numbers.


\begin{acknowledgments}
Work supported by:
EU Horizon2020 research and innovation program, STRONG-2020project, under grant agreement no. 824093;
Ministerio Espa\~nol de Ciencia e Innovaci\'on, grant no. PID2019-107844GB-C22 and PID2019-105439GB-C22/AEI/10.13039/501100011033;
and Junta de Andaluc\'ia, contract nos.\ P18-FR-5057 and Operativo FEDER Andaluc\'ia 2014-2020 UHU-1264517.
\end{acknowledgments}



\bibliography{CQM_Zbs}

\end{document}